\documentclass[cameraready]{Interspeech}

\title{MeanVC 2: Robust Low-Latency Streaming Zero-Shot Voice Conversion}

\author[affiliation={1}, equalcontribution]{Guobin}{Ma}
\author[affiliation={1}, equalcontribution]{Yuxuan}{Xia}
\author[affiliation={1},]{Yuepeng}{Jiang}
\author[affiliation={1},]{Dake}{Guo}
\author[affiliation={1},]{Hanke}{Xie}
\author[affiliation={1},]{Jingbin}{Hu}
\author[affiliation={2},]{Yanbo}{Wang}
\author[affiliation={1},correspondingauthor]{Lei}{Xie}
\author[affiliation={3},correspondingauthor]{Pengcheng}{Zhu}

\address{
$^{1}$ Audio, Speech and Language Processing Group (ASLP@NPU), School of Software, \\
Northwestern Polytechnical University, China \\
$^{2}$ The University of New South Wales, Australia  \\
$^{3}$ WeNet Open Source Community, China
}
\email{
guobin.ma@mail.nwpu.edu.cn,
lxie@nwpu.edu.cn, 
zpcoftts@gmail.com
}

\keywords{streaming voice conversion, zero-shot, mean flows, universal timbre token encoder}

\usepackage{comment}
\usepackage{makecell}
\usepackage{amsmath}
\usepackage{multirow}
\usepackage[table]{xcolor}
\definecolor{lightgreen}{RGB}{230,245,230}

\begin{document}

\maketitle

\begin{abstract}
   Streaming zero-shot voice conversion (VC) has become increasingly popular due to its potential for real-time applications. 
   The recently proposed MeanVC achieves lightweight streaming zero-shot VC, but it has several limitations: its chunk-wise autoregressive denoising doubles the effective training sequence length, conversion quality degrades under small-chunk settings, and its timbre encoder directly relies on reference mel-spectrograms, making it sensitive to reference audio quality. To address these limitations we propose MeanVC~2. We introduce future-receptive chunking (FRC), which explicitly schedules past and future receptive fields across diffusion transformer decoder layers and removes clean-chunk teacher forcing. By incorporating bounded future context, FRC enables stable conversion with a 40~ms chunk size. 
   We further introduce a universal timbre token encoder, which constructs a timbre representation from a global speaker embedding and retrieves fine-grained timbre cues via cross-attention, improving robustness to low-quality references and enhancing zero-shot speaker similarity. 
   Experimental results show that MeanVC~2 significantly outperforms MeanVC, while reducing latency from 211\,ms to 110\,ms.
   Audio samples are publicly available\footnote{\url{https://aslp-lab.github.io/MeanVC2/}}. 
   The source code will be publicly released.
\end{abstract}

\section{Introduction}

Zero-shot voice conversion (VC) aims to transform the timbre of a source speaker into that of an arbitrary unseen target speaker while preserving the underlying linguistic content~\cite{DBLP:journals/taslp/SismanYKL21}. 
This technology enables diverse practical applications, including movie dubbing~\cite{bgmvc, expressive-vc}, privacy protection~\cite{DBLP:conf/icassp/YaoWLGXWL23}, and communication aids for individuals with speech impairments~\cite{ELECTROLARYNGEAL}. 
Driven by continuous advances in deep learning, recent models~\cite{DBLP:conf/icassp/HussainNHLG23, DBLP:conf/icassp/LiLL23, DBLP:conf/icassp/SEF-VC, DBLP:conf/icassp/LuoD24,DBLP:conf/icassp/KimKCNMC25,DBLP:conf/icassp/ChoiP25,jiang2025ref} demonstrate substantial improvements in generation naturalness and speaker similarity. 
Concurrent with these developments, voice conversion gains increasing traction in real-time communication (RTC) scenarios, such as live broadcasting, online meetings, and voice chat in multiplayer games. 
This widespread adoption highlights a pressing need for streaming models that deliver low latency and low computational cost while maintaining high fidelity in both audio quality and speaker similarity.

\begin{figure*}[t]
  \centering
  \includegraphics[width=\textwidth]{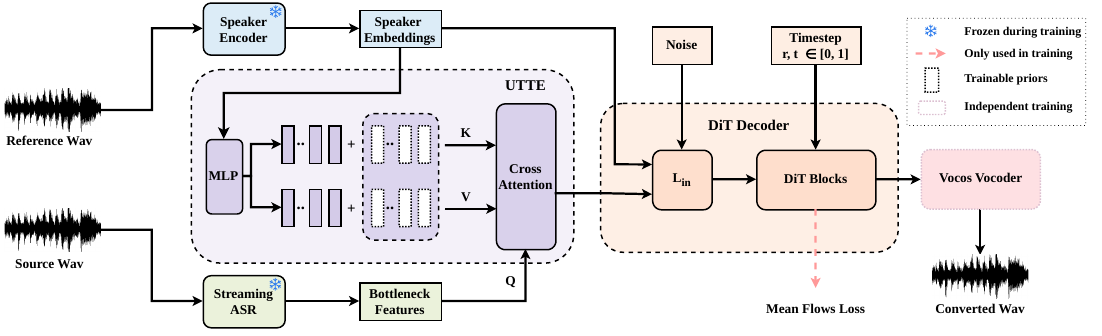}
  \caption{Overall architecture of our proposed MeanVC~2.}
  \label{fig:overview}
  \vspace{-13pt}
\end{figure*}

Recent streaming zero-shot VC methods primarily follow two architectural paradigms: autoregressive (AR) and non-autoregressive (NAR) frameworks.
AR methods~\cite{DBLP:conf/acl/WangCW0W24,DBLP:journals/spl/WangCWXW24} typically employ strictly causal, step-by-step generation to avoid future look-ahead. While this often yields strong fidelity and stable speaker characteristics, sequential decoding inevitably introduces substantial latency and computational overhead, hindering lightweight real-time deployment.
In contrast, NAR methods~\cite{DBLP:conf/icassp/NingJ0WY0B24,DBLP:journals/corr/seedvc} reduce latency through parallel or few-step generation, making them more suitable for streaming scenarios. However, under a limited streaming context, particularly during short-chunk processing, these methods are prone to degraded intelligibility, audio quality, speaker similarity, and cross-chunk consistency.
Consequently, achieving low latency and low computational cost without sacrificing fidelity and robustness remains a central challenge in streaming zero-shot VC.

MeanVC~\cite{DBLP:journals/corr/abs-2510-08392} has attempted to address these challenges by combining chunk-wise autoregressive denoising (CARD) with mean flows~\cite{DBLP:journals/corr/MeanFlows}. 
Specifically, CARD enables streaming conversion through a chunk-wise causal attention mask, in which each noisy chunk attends to a limited window of previously generated clean chunks, thereby preserving cross-chunk consistency. In addition, mean flows improves the efficiency of flow matching by training the model to regress the average velocity field along the ordinary differential equation (ODE) trajectory, enabling high-quality spectrogram synthesis with only a single neural function evaluation (1-NFE).
Despite its effectiveness, MeanVC still has several limitations that constrain its practical performance. 
First, the CARD training scheme is training-unfriendly because it doubles the effective sequence length, increases memory consumption, and converges slowly in practice.
Second, with a 160\,ms chunk size, MeanVC has an end-to-end first-packet latency of 211\,ms, and its performance degrades noticeably with smaller chunk sizes.
Third, MeanVC relies on a multi-reference timbre encoder (MRTE)~\cite{DBLP:conf/iclr/0001L0HYJY0WW0M24} that extracts fine-grained speaker characteristics directly from reference mel-spectrograms, which makes speaker conditioning sensitive to reference audio quality and consequently leads to degraded speaker similarity under low-quality references.

To address these limitations we propose \textbf{MeanVC~2}, a streaming zero-shot voice conversion system that delivers enhanced robustness and reduced latency.
Specifically, we propose \textit{future-receptive chunking} (FRC), a block-wise receptive-field scheduling strategy for a decoder based on the diffusion transformer (DiT)~\cite{DBLP:conf/iccv/PeeblesX23}, which assigns a dedicated attention mask at each layer to explicitly control how each chunk attends to past and future context. 
Compared with the CARD-based design in MeanVC, FRC eliminates the need for clean-chunk teacher forcing, thereby substantially reducing training memory consumption and improving overall training efficiency.
Furthermore, by incorporating bounded future context into the receptive field, FRC alleviates the acoustic context insufficiency that causes quality degradation under short chunks, enabling stable conversion with a 40~ms chunk size.
Finally, we propose the \textit{universal timbre token encoder} (UTTE) to improve the robustness of speaker conditioning under low-quality reference audio. 
UTTE decouples fine-grained timbre extraction from direct reliance on reference mel-spectrograms by constructing a key--value timbre representation parameterized by universal timbre tokens (UTT) from a global speaker embedding, and using bottleneck features (BNFs) as queries to retrieve pronunciation-aware timbre details via cross-attention.
This design improves robustness to low-quality references, enhances zero-shot speaker similarity, and improves data scalability by reducing dependence on large-scale speaker-labeled reference mel data.
Extensive experiments demonstrate that MeanVC~2 substantially outperforms MeanVC in overall performance, reducing the end-to-end pipeline latency on a single CPU core from 211\,ms to \textbf{110\,ms}. The source code will be publicly released.

\section{Preliminaries}
\subsection{Conditional flow matching}

Conditional flow matching (CFM)~\cite{DBLP:conf/iclr/LipmanCBNL23} learns a vector field to transport samples from a prior distribution $p_{\text{prior}}(\epsilon)$ to a data distribution $p_{\text{data}}(x)$. Given a data sample $x \sim p_{\text{data}}(x)$ and noise $\epsilon \sim \mathcal{N}(0, I)$, an optimal transport path is constructed as $z_t = (1-t)x + t\epsilon$, with the conditional velocity $v_t = dz_t/dt = \epsilon - x$. A neural network $f_\theta$ is trained to minimize the CFM objective:
\begin{equation}
\mathcal{L}_{\text{CFM}}(\theta)
= \mathbb{E}_{t,x,\epsilon}\left[
\left\lVert f_{\theta}(t,z_t)-v_t \right\rVert^2
\right].
\end{equation}
During inference, the model solves an ODE to recover $x$ from $\epsilon$, which typically requires multiple function evaluations and incurs non-trivial latency.

\subsection{Mean flows}

To enable high-quality synthesis with only a single neural function evaluation (1-NFE), this work adopts mean flows~\cite{DBLP:journals/corr/MeanFlows}. Given a time interval $[r, t]$, the average velocity along the ODE trajectory is defined as:
\begin{equation}
u(z_t, r, t) \triangleq \frac{1}{t-r} \int_r^t v(z_\tau, \tau)\, d\tau.
\end{equation}
Differentiating with respect to $t$ and rearranging yields the mean flows identity:
\begin{equation}
u(z_t, r, t) = v(z_t, t) - (t - r)\frac{d}{dt}u(z_t, r, t),
\end{equation}
where the total derivative is expanded via the Jacobian-vector product as $\frac{d}{dt}u = v(z_t, t)\partial_z u + \partial_t u$. Replacing the marginal velocity $v(z_t, t)$ with the conditional velocity $v_t = \epsilon - x$, the training target becomes $u_{\text{tgt}} = v_t - (t-r)(v_t \partial_z u_\theta + \partial_t u_\theta)$. The mean flows training objective is:
\begin{equation}
\mathcal{L}_{\text{MF}}(\theta)
= \mathbb{E}_{t,r,x,\epsilon}\left[
\left\lVert f_\theta(z_t,r,t)-\operatorname{sg}(u_{\text{tgt}}) \right\rVert^2
\right].
\end{equation}
where $\text{sg}(\cdot)$ denotes the stop-gradient operation. At $t = r$, this objective reduces to the standard CFM loss. During 1-NFE sampling, the clean sample is recovered as $z_0 = z_1 - f_\theta(z_1, 0, 1)$, where $z_1 = \epsilon \sim p_{\text{prior}}(\epsilon)$.

\section{Methods}

\subsection{Overview}

As illustrated in Fig.~\ref{fig:overview}, MeanVC~2 follows a recognition--synthesis framework and consists of a streaming automatic speech recognition (ASR) module, a speaker encoder, a \textit{universal timbre token encoder} (UTTE), a DiT decoder, and a vocoder. First, a pretrained streaming ASR model extracts bottleneck features (BNFs) from the source waveform, while a pretrained speaker encoder extracts a global speaker embedding from the reference waveform. The BNFs and the global speaker embedding are then fed into UTTE. Specifically, UTTE transforms the global speaker embedding into a UTT-based key--value timbre representation and uses the BNFs as queries in a cross-attention module to retrieve fine-grained, content-aware timbre cues, thereby producing timbre-aware BNFs.

Conditioned on the timbre-aware BNFs, the DiT decoder generates the target mel-spectrogram in a streaming manner using the proposed FRC strategy. Following MeanVC, the decoder is trained with the mean flows formulation, which enables high-quality mel-spectrogram synthesis with 1-NFE while preserving the efficiency advantage of the original framework. 

\subsection{Future-receptive chunking}

MeanVC adopts chunk-wise autoregressive denoising, where clean mel chunks serve as prompts to condition noisy chunks. This design preserves cross-chunk consistency but requires concatenating all clean chunks and all noisy chunks into a $2N$-chunk sequence during training, which substantially increases memory cost and makes optimization less training-friendly. To improve training efficiency, we instead perform chunked training with a block-wise attention mask on the noisy sequence, avoiding clean-chunk teacher forcing. In our implementation, under the same training configuration, this modification reduces the peak GPU memory footprint by about 60\%. However, restricting attention to a short history window tends to fragment temporal context into small chunks, degrading cross-chunk consistency and audio quality.

To mitigate these limitations, we introduce FRC, a block-wise receptive-field scheduling strategy for the DiT-based decoder. 
FRC partitions the temporal sequence into $N$ chunks ${C_i}_{i=1}^{N}$, each containing $B$ consecutive frames, and assigns a dedicated attention mask at each DiT layer to explicitly control how chunk $C_i$ attends to preceding and subsequent chunks. Specifically, FRC uses three complementary mask types: an intra-chunk mask that permits full attention within each chunk, a backward mask that allows $C_i$ to attend to a limited number of preceding chunks, and a forward mask that allows it to attend to a limited number of subsequent chunks. Through this layer-wise scheduling, the receptive field expands across successive DiT layers in a controlled manner, rather than relying on a fixed sliding-window design.

\begin{figure}[h]
  \centering
  \includegraphics[width=0.85\linewidth]{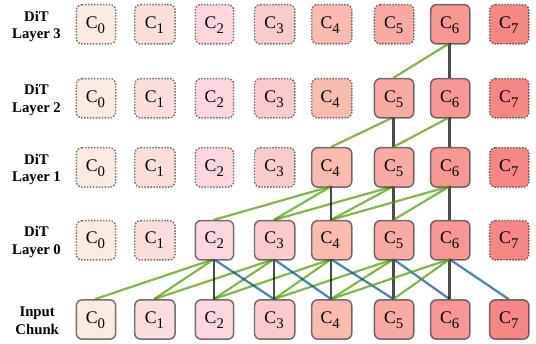}
  \caption{Layer-wise receptive-field expansion of chunk $\text{C}_6$ under FRC in a 4-layer DiT decoder. Green, black, and blue edges denote dependencies introduced by the backward, intra-chunk, and forward masks, respectively.}
  \label{fig:FRC_Receptive_Field}
  \vspace{-13pt}
\end{figure}

Concretely, for a decoder with $L$ layers, each chunk at layer $\ell$ can attend to at most $P_\ell$ past chunks and $F_\ell$ future chunks, while full intra-chunk attention is always preserved. This design defines a block mask $M^{(\ell)}$ at layer $\ell$, where attention from chunk $i$ to chunk $j$ is allowed only if $j \in [i-P_\ell, i+F_\ell]$. In our implementation with $L=4$, the past receptive fields are set to ${P_\ell}={2,2,1,1}$ and the future receptive fields are set to ${F_\ell}={1,0,0,0}$, as shown in Fig.~\ref{fig:FRC_Receptive_Field}. As a result, the overall receptive field covers $6$ past chunks, the current chunk, and $1$ future chunk, for a total of $8$ chunks. This explicit receptive-field scheduling with bounded look-ahead alleviates the lack of acoustic context in short-chunk settings and provides a more stable temporal context for streaming generation.

\subsection{Universal timbre token encoder}

MeanVC relies on MRTE, which extracts fine-grained speaker characteristics directly from reference mel-spectrograms of the same speaker. 
This design makes speaker conditioning sensitive to the acoustic quality of the reference audio at inference time, often reducing speaker similarity when the reference is degraded. It also limits data scalability, since collecting large-scale reference audio with reliable speaker identity labels is difficult. Inspired by TVTSYN~\cite{quamer2026tvtsyn}, we propose UTTE to decouple fine-grained timbre extraction from direct reliance on reference mel-spectrograms.

UTTE first maps the global speaker embedding $s$, extracted by the speaker encoder, into a set of UTT parameterized as $K$ key--value pairs $\{(k_i, v_i)\}_{i=1}^{K}$. The UTT adopts a dual representation consisting of a speaker-specific component generated from $s$ by a multilayer perceptron (MLP) and a set of learnable priors shared across all speakers, denoted by $k_i^{\text{prior}}$ and $v_i^{\text{prior}}$. Each key-value pair is computed as:

\begin{equation}
\begin{aligned}
k_i &= \mathrm{MLP}_k(s)_i + \tanh(k_i^{\text{prior}}), \\
v_i &= \mathrm{MLP}_v(s)_i + \tanh(v_i^{\text{prior}}).
\end{aligned}
\end{equation}
The learnable priors act as universal timbre prototypes that capture speaker-agnostic phonation traits such as breathiness and nasality, while the MLP outputs modulate these prototypes to reflect the identity of the target speaker. Applying $\tanh(\cdot)$ to the prior terms before additive fusion empirically improves token diversity. Intuitively, the UTT decomposes timbre into multiple facets each stored in a dedicated slot, which provides a strong inductive bias and improves both sample efficiency and training stability.

Unlike TVTSYN, which applies scaled dot-product attention between content embeddings and the UTT and then combines the resulting features with a global timbre embedding via spherical linear interpolation to obtain a time-varying timbre embedding, UTTE directly uses BNFs as queries to attend over the UTT key-value pairs, thereby extracting fine-grained, pronunciation-aware timbre features and producing timbre-aware BNFs for downstream spectrogram generation.

\begin{table*}[ht]
  \centering
  \caption{Zero-shot voice conversion results. RTF is measured for the VC module only, while latency denotes the end-to-end first-packet latency of the full pipeline. The best and second-best results are highlighted in bold and underlined, respectively, among the main systems (excluding ablations). Rows shaded in light green correspond to MeanVC~2 and its ablations.}
  \vspace{-7pt}
  \label{tab:zeroshot}
  \small
  \renewcommand\arraystretch{1.22}
  \resizebox{0.95\textwidth}{!}
  {
    \begin{tabular}{ l c c c c c c c c c c}
    \toprule
    \multirow{2}{*}{Method}
      & \multicolumn{3}{c}{Quality} 
      & & \multicolumn{2}{c}{Similarity}
      & & \multicolumn{3}{c}{Efficiency}
      \\ \cline{2-4} \cline{6-7} \cline{9-11}
        & NMOS$\uparrow$ & DNSMOS$\uparrow$ & CER(\%)$\downarrow$ & & SMOS$\uparrow$ & SSIM$\uparrow$ & & Parameters(M)  & RTF$\downarrow$   & Latency(ms)$\downarrow$ \\ 
      \midrule
      GT                        & 4.07$\pm$0.02& 3.79  & 1.36     &   &  -    & -   &   & -  & - & -     \\
      \midrule
      StreamVoice+& 3.70$\pm$0.04& 3.52& 10.27&   &  3.65$\pm$0.02& 0.552&  & 153& 14.732& 1258.56\\
      
      MeanVC~(80\,ms)             & 3.61$\pm$0.02& 3.37& 11.66&   &  3.61$\pm$0.03& 0.599&  & 14& \underline{0.177}& \underline{111.64}\\

      MeanVC~(160\,ms)            & \textbf{3.86$\pm$0.04}& \underline{3.81}& \textbf{5.11}&   &  \underline{3.87$\pm$0.03}& \underline{0.687}&  & 14& \textbf{0.136}& 211.52\\
      \rowcolor{lightgreen}
      MeanVC~2                  & \underline{3.81$\pm$0.05}& \textbf{3.89}& \underline{7.44}&   &  \textbf{3.89$\pm$0.04}& \textbf{0.710}&  & 18& 0.371& \textbf{109.88}\\
      \midrule
      \rowcolor{lightgreen}
      \quad w/o forward mask    & 3.54$\pm$0.02& 3.23& 20.65&   &  3.52$\pm$0.02& 0.573&  & 18& -& -\\
      \rowcolor{lightgreen}
      \quad w/o UTTE            & 3.77$\pm$0.05& 3.81& 7.92&   &  3.78$\pm$0.02& 0.682&  & 13& -& -\\
      \rowcolor{lightgreen}
      \quad w/o tanh            & 3.79$\pm$0.03& 3.83& 7.79&   &  3.82$\pm$0.05& 0.692&  & 18& -& -\\
      
      \bottomrule
    \end{tabular}
    }
\vspace{-15pt}
\end{table*}

\section{Experiments}
\subsection{Experiment setup}
\noindent\textbf{Dataset.}
We train MeanVC~2 on the open-source Emilia~\cite{DBLP:conf/slt/Emilia} corpus. Specifically, we filter out utterances shorter than 5\,s and randomly sample 10{,}000 hours of Mandarin data. All audio files are resampled to 16\,kHz for VC training. 
For zero-shot evaluation, we use the Mandarin subset of the Seed-TTS test set~\cite{DBLP:journals/corr/Seed-tts}, comprising 2{,}018 source--target test pairs. 
To evaluate the robustness of speaker conditioning under low-quality reference audio, we further select 30 target speakers whose reference recordings exhibit poor acoustic quality and sample 100 utterances from the Seed-TTS test set as source recordings.
To extract semantic features, we incorporate the streaming ASR model Fast-U2++~\cite{DBLP:conf/icassp/Fast-U2++}, implemented by WeNet~\cite{DBLP:conf/interspeech/WeNet} and trained on WenetSpeech~\cite{DBLP:conf/icassp/WENETSPEECH}.

\noindent\textbf{Implementation details.}
MeanVC~2 contains 18M parameters in total. The DiT decoder uses four DiT blocks, each with a hidden size of 512 and two attention heads. 
The UTTE includes two independent two-layer MLPs, 32 UTT key-value pairs with a hidden size of 256, and a cross-attention module with a hidden size of 256 and four attention heads.
Fast-U2++ uses an 80\,ms chunk size for streaming inference and extracts semantic features from 16\,kHz waveforms with a 40\,ms frame length.
Speaker embeddings are extracted using ECAPA-TDNN~\cite{DBLP:conf/interspeech/ECAPA-TDNN}.
We employ Vocos~\cite{DBLP:conf/iclr/Vocos} as the vocoder to convert mel-spectrograms to high-fidelity 16\,kHz speech waveforms.

\noindent\textbf{Evaluation metrics.}
For subjective evaluation, we use naturalness mean opinion score (NMOS) and speaker similarity mean opinion score (SMOS), each reported with 95\% confidence intervals. We randomly sample 100 test pairs and recruit 30 listeners for the listening tests. 
For objective evaluation, we measure intelligibility using character error rate (CER) computed by an ASR model\footnote{https://huggingface.co/funasr/paraformer-zh}. We measure speaker similarity (SSIM) using a WavLM-finetuned speaker verification model\footnote{https://github.com/BytedanceSpeech/seed-tts-eval} to compare converted speech with the target speaker. In addition, we use DNSMOS\footnote{https://github.com/microsoft/DNS-Challenge} to assess speech quality. For streaming performance, we report real-time factor (RTF) and latency. RTF is defined as the ratio between inference time and the generated audio duration. All RTF and latency measurements are conducted with single-threaded execution on a single-core AMD EPYC 7542 CPU.

\noindent\textbf{Baseline systems.}
To assess the effectiveness of the proposed approach, we evaluate MeanVC~2 against two baseline systems, MeanVC and StreamVoice+. MeanVC~2 uses 40\,ms chunks and introduces a bounded future receptive field of one chunk, resulting in an effective first-packet buffering of 80\,ms at the BNF level. StreamVoice+ uses an 80\,ms chunk size. MeanVC is evaluated with chunk sizes of 80\,ms and 160\,ms.

\subsection{Evaluation on zero-shot VC}

As shown in Table~\ref{tab:zeroshot}, MeanVC~2 attains the highest scores across all similarity metrics, surpassing StreamVoice+, MeanVC (80\,ms), and MeanVC (160\,ms). In terms of quality metrics, MeanVC~2 achieves the highest DNSMOS~\cite{DBLP:conf/icassp/Dnsmos}.
Since MeanVC~2 and MeanVC (80\,ms) use the same 80\,ms input chunk, this comparison is fair, and MeanVC~2 improves all five quality and similarity metrics over MeanVC (80\,ms).
The slight disadvantage in NMOS and CER relative to MeanVC (160\,ms) is expected, as the 160\,ms setting benefits from richer contextual information for acoustic reconstruction.

MeanVC~2 contains only 18M parameters, comparable to MeanVC at 14M and much smaller than StreamVoice+ at 153M. Its end-to-end first-packet latency is 109.88\,ms, comparable to MeanVC (80\,ms) at 111.64\,ms, but much lower than MeanVC (160\,ms) at 211.52\,ms and StreamVoice+ at 1258.56\,ms. The full pipeline remains real-time, with an overall RTF of $0.114 + 0.371 + 0.148 = 0.633 < 1.0$ for the ASR encoder, VC module, and vocoder. Although the VC-module RTF of MeanVC~2 is higher than that of MeanVC (80\,ms), i.e., 0.371 versus 0.177, the latter is a latency-matched baseline rather than an output-duration-matched computation reference. Under the same 40\,ms output granularity, MeanVC has an RTF of 0.316, compared with 0.371 for MeanVC~2, indicating that MeanVC~2 only moderately increases the computational cost. All efficiency measurements were conducted without engineering optimizations such as quantization or operator fusion, suggesting that further latency reduction remains possible.

\vspace{-2pt}
\subsection{Evaluation on reference robustness}

As shown in Table~\ref{tab:robustness}, replacing UTTE with MRTE leads to consistent degradation across all metrics. These results demonstrate that UTTE extracts more reliable speaker representations from low-quality reference audio than MRTE, resulting in better output quality and greater speaker similarity under such conditions.

\begin{table}[ht]
  \centering
  \caption{Evaluation results on reference robustness.}
  \vspace{-2pt}
  \label{tab:robustness}
  \fontsize{8}{10}\selectfont 
  \renewcommand\arraystretch{1.12}
  \resizebox{0.45\textwidth}{!}{
    \begin{tabular}{ l c c c}
    \toprule
    Method                   & DNSMOS$\uparrow$ & CER(\%)$\downarrow$ & SSIM$\uparrow$ \\ 
    \midrule
    MeanVC~2 w/ MRTE    & 1.39             & 7.64                & 0.621          \\
    MeanVC~2                 & \textbf{1.87}    & \textbf{6.55}       & \textbf{0.643} \\
    \bottomrule
    \end{tabular}
  }
\vspace{-4pt}
\end{table}

\vspace{-6pt}
\subsection{Ablation study}

To validate the contribution of each component, we conduct ablation experiments as shown in Table~\ref{tab:zeroshot}. Removing the forward mask from FRC results in the most severe degradation, indicating that a limited lookahead receptive field is essential for capturing sufficient acoustic context under small chunk sizes. Removing UTTE leads to a notable drop in SSIM, confirming that UTTE captures fine-grained timbre information beyond what a global speaker embedding alone can provide; notably, even this reduced 13M-parameter variant still outperforms MeanVC (80\,ms) across all metrics, highlighting the effectiveness of FRC under small chunk sizes.
 Removing the tanh activation causes a moderate drop in similarity metrics, suggesting that tanh regularizes the distribution of utterance-level timbre embeddings and promotes greater diversity in speaker representations.

\section{Conclusion}

We propose MeanVC~2, a low-latency and robust streaming zero-shot VC system that addresses key limitations of MeanVC through FRC and UTTE. FRC removes clean-chunk teacher forcing to improve training efficiency and incorporates bounded future context to stabilize short-chunk conversion, enabling reliable conversion with a 40~ms chunk size. UTTE improves speaker conditioning robustness by retrieving fine-grained timbre cues from a global speaker embedding, reducing sensitivity to low-quality reference audio. 
Experimental results demonstrate that MeanVC~2 substantially outperforms MeanVC in overall performance, reducing end-to-end pipeline latency from 211\,ms to 110\,ms.

\newpage

\section{Generative AI Use Disclosure}
In accordance with ISCA guidelines, the authors declare that all intellectual contributions to this manuscript---including core ideas, theoretical formulation, methodology, experimental design, result analysis, and conclusions---originate entirely from the authors. Generative AI was used solely to assist with language editing and writing fluency. The authors take full responsibility for the accuracy, integrity, and originality of this work. No generative AI tool is listed as a co-author, and all authors have reviewed and approved the final submission.

\bibliographystyle{IEEEtran}
\bibliography{mybib}

@inproceedings{DBLP:conf/icassp/HussainNHLG23,
  author       = {Shehzeen Hussain and
                  Paarth Neekhara and
                  Jocelyn Huang and
                  Jason Li and
                  Boris Ginsburg},
  title        = {{ACE-VC:} Adaptive and Controllable Voice Conversion Using Explicitly
                  Disentangled Self-Supervised Speech Representations},
  booktitle    = {{ICASSP}},
  pages        = {1--5},
  publisher    = {{IEEE}},
  year         = {2023}
}

@inproceedings{DBLP:conf/icassp/LiLL23,
  author       = {Dayong Li and
                  Xian Li and
                  Xiaofei Li},
  title        = {{DVQVC:} An Unsupervised Zero-Shot Voice Conversion Framework},
  booktitle    = {{ICASSP}},
  pages        = {1--5},
  publisher    = {{IEEE}},
  year         = {2023}
}

@inproceedings{DBLP:conf/icassp/SEF-VC,
  author       = {Junjie Li and
                  Yiwei Guo and
                  Xie Chen and
                  Kai Yu},
  title        = {{SEF-VC:} Speaker Embedding Free Zero-Shot Voice Conversion with Cross
                  Attention},
  booktitle    = {{ICASSP}},
  pages        = {12296--12300},
  publisher    = {{IEEE}},
  year         = {2024}
}

@inproceedings{DBLP:conf/icassp/LuoD24,
  author       = {Yin{-}Jyun Luo and
                  Simon Dixon},
  title        = {Posterior Variance-Parameterised Gaussian Dropout: Improving Disentangled
                  Sequential Autoencoders for Zero-Shot Voice Conversion},
  booktitle    = {{ICASSP}},
  pages        = {11676--11680},
  publisher    = {{IEEE}},
  year         = {2024}
}

@inproceedings{DBLP:conf/icassp/KimKCNMC25,
  author       = {Jaehun Kim and
                  Ji{-}Hoon Kim and
                  Yeunju Choi and
                  Tan Dat Nguyen and
                  Seongkyu Mun and
                  Joon Son Chung},
  title        = {AdaptVC: High Quality Voice Conversion with Adaptive Learning},
  booktitle    = {{ICASSP}},
  pages        = {1--5},
  publisher    = {{IEEE}},
  year         = {2025}
}

@inproceedings{DBLP:conf/icassp/ChoiP25,
  author       = {Ha{-}Yeong Choi and
                  Jaehan Park},
  title        = {VoicePrompter: Robust Zero-Shot Voice Conversion with Voice Prompt
                  and Conditional Flow Matching},
  booktitle    = {{ICASSP}},
  pages        = {1--5},
  publisher    = {{IEEE}},
  year         = {2025}
}

@inproceedings{DBLP:conf/acl/WangCW0W24,
  author       = {Zhichao Wang and
                  Yuanzhe Chen and
                  Xinsheng Wang and
                  Lei Xie and
                  Yuping Wang},
  title        = {StreamVoice: Streamable Context-Aware Language Modeling for Real-time
                  Zero-Shot Voice Conversion},
  booktitle    = {{ACL} {(1)}},
  pages        = {7328--7338},
  publisher    = {Association for Computational Linguistics},
  year         = {2024}
}

@inproceedings{DBLP:conf/icassp/NingJ0WY0B24,
  author       = {Ziqian Ning and
                  Yuepeng Jiang and
                  Pengcheng Zhu and
                  Shuai Wang and
                  Jixun Yao and
                  Lei Xie and
                  Mengxiao Bi},
  title        = {Dualvc 2: Dynamic Masked Convolution for Unified Streaming and Non-Streaming
                  Voice Conversion},
  booktitle    = {{ICASSP}},
  pages        = {11106--11110},
  publisher    = {{IEEE}},
  year         = {2024}
}

@inproceedings{DBLP:conf/iccv/PeeblesX23,
  author       = {William Peebles and
                  Saining Xie},
  title        = {Scalable Diffusion Models with Transformers},
  booktitle    = {{ICCV}},
  pages        = {4172--4182},
  publisher    = {{IEEE}},
  year         = {2023}
}

@inproceedings{DBLP:conf/iclr/LipmanCBNL23,
  author       = {Yaron Lipman and
                  Ricky T. Q. Chen and
                  Heli Ben{-}Hamu and
                  Maximilian Nickel and
                  Matthew Le},
  title        = {Flow Matching for Generative Modeling},
  booktitle    = {{ICLR}},
  publisher    = {OpenReview.net},
  year         = {2023}
}

@inproceedings{DBLP:conf/slt/Emilia,
  author       = {Haorui He and
                  Zengqiang Shang and
                  Chaoren Wang and
                  Xuyuan Li and
                  Yicheng Gu and
                  Hua Hua and
                  Liwei Liu and
                  Chen Yang and
                  Jiaqi Li and
                  Peiyang Shi and
                  Yuancheng Wang and
                  Kai Chen and
                  Pengyuan Zhang and
                  Zhizheng Wu},
  title        = {Emilia: An Extensive, Multilingual, and Diverse Speech Dataset For
                  Large-Scale Speech Generation},
  booktitle    = {{SLT}},
  pages        = {885--890},
  publisher    = {{IEEE}},
  year         = {2024}
}

@inproceedings{DBLP:conf/icassp/Dnsmos,
  author       = {Chandan K. A. Reddy and
                  Vishak Gopal and
                  Ross Cutler},
  title        = {Dnsmos {P.835:} {A} Non-Intrusive Perceptual Objective Speech Quality
                  Metric to Evaluate Noise Suppressors},
  booktitle    = {{ICASSP}},
  pages        = {886--890},
  publisher    = {{IEEE}},
  year         = {2022}
}

@inproceedings{DBLP:conf/icassp/Fast-U2++,
  author       = {Chengdong Liang and
                  Xiao{-}Lei Zhang and
                  Binbin Zhang and
                  Di Wu and
                  Shengqiang Li and
                  Xingchen Song and
                  Zhendong Peng and
                  Fuping Pan},
  title        = {Fast-U2++: Fast and Accurate End-to-End Speech Recognition in Joint
                  CTC/Attention Frames},
  booktitle    = {{ICASSP}},
  pages        = {1--5},
  publisher    = {{IEEE}},
  year         = {2023}
}

@inproceedings{DBLP:conf/interspeech/WeNet,
  author       = {Zhuoyuan Yao and
                  Di Wu and
                  Xiong Wang and
                  Binbin Zhang and
                  Fan Yu and
                  Chao Yang and
                  Zhendong Peng and
                  Xiaoyu Chen and
                  Lei Xie and
                  Xin Lei},
  title        = {WeNet: Production Oriented Streaming and Non-Streaming End-to-End
                  Speech Recognition Toolkit},
  booktitle    = {Interspeech},
  pages        = {4054--4058},
  publisher    = {{ISCA}},
  year         = {2021}
}

@inproceedings{DBLP:conf/icassp/WENETSPEECH,
  author       = {Binbin Zhang and
                  Hang Lv and
                  Pengcheng Guo and
                  Qijie Shao and
                  Chao Yang and
                  Lei Xie and
                  Xin Xu and
                  Hui Bu and
                  Xiaoyu Chen and
                  Chenchen Zeng and
                  Di Wu and
                  Zhendong Peng},
  title        = {{WENETSPEECH:} {A} 10000+ Hours Multi-Domain Mandarin Corpus for Speech
                  Recognition},
  booktitle    = {{ICASSP}},
  pages        = {6182--6186},
  publisher    = {{IEEE}},
  year         = {2022}
}

@inproceedings{DBLP:conf/interspeech/ECAPA-TDNN,
  author       = {Brecht Desplanques and
                  Jenthe Thienpondt and
                  Kris Demuynck},
  title        = {{ECAPA-TDNN:} Emphasized Channel Attention, Propagation and Aggregation
                  in {TDNN} Based Speaker Verification},
  booktitle    = {{INTERSPEECH}},
  pages        = {3830--3834},
  publisher    = {{ISCA}},
  year         = {2020}
}

@inproceedings{DBLP:conf/iclr/Vocos,
  author       = {Hubert Siuzdak},
  title        = {Vocos: Closing the gap between time-domain and Fourier-based neural
                  vocoders for high-quality audio synthesis},
  booktitle    = {{ICLR}},
  publisher    = {OpenReview.net},
  year         = {2024}
}

@inproceedings{bgmvc,
  author       = {Jixun Yao and
                  Yi Lei and
                  Qing Wang and
                  Pengcheng Guo and
                  Ziqian Ning and
                  Lei Xie and
                  Hai Li and
                  Junhui Liu and
                  Danming Xie},
  title        = {Preserving background sound in noise-robust voice conversion via multi-task
                  learning},
  booktitle    = {Proc. ICASSP},
  pages        = {1--5},
  publisher    = {{IEEE}},
  year         = {2023},
}

@inproceedings{expressive-vc,
  author       = {Ziqian Ning and
                  Qicong Xie and
                  Pengcheng Zhu and
                  Zhichao Wang and
                  Liumeng Xue and
                  Jixun Yao and
                  Lei Xie and
                  Mengxiao Bi},
  title        = {Expressive-VC: Highly Expressive Voice Conversion with Attention Fusion
                  of Bottleneck and Perturbation Features},
  booktitle    = {Proc. ICASSP},
  pages        = {1--5},
  publisher    = {{IEEE}},
  year         = {2023},
}

@INPROCEEDINGS{ELECTROLARYNGEAL,
  author={Kobayashi, Kazuhiro and Hayashi, Tomoki and Toda, Tomoki},
  booktitle={Proc. ICASSP}, 
  title={Low-Latency Electrolaryngeal Speech Enhancement Based on Fastspeech2-Based Voice Conversion and Self-Supervised Speech Representation}, 
  year={2023},
  volume={},
  number={},
  publisher    = {{IEEE}},
  pages={1-5},
}

@inproceedings{DBLP:conf/icassp/YaoWLGXWL23,
  author       = {Jixun Yao and
                  Qing Wang and
                  Yi Lei and
                  Pengcheng Guo and
                  Lei Xie and
                  Namin Wang and
                  Jie Liu},
  title        = {Distinguishable Speaker Anonymization Based on Formant and Fundamental
                  Frequency Scaling},
  booktitle    = {{ICASSP}},
  pages        = {1--5},
  publisher    = {{IEEE}},
  year         = {2023}
}

@inproceedings{DBLP:conf/iclr/0001L0HYJY0WW0M24,
  author       = {Ziyue Jiang and
                  Jinglin Liu and
                  Yi Ren and
                  Jinzheng He and
                  Zhenhui Ye and
                  Shengpeng Ji and
                  Qian Yang and
                  Chen Zhang and
                  Pengfei Wei and
                  Chunfeng Wang and
                  Xiang Yin and
                  Zejun Ma and
                  Zhou Zhao},
  title        = {Mega-TTS 2: Boosting Prompting Mechanisms for Zero-Shot Speech Synthesis},
  booktitle    = {{ICLR}},
  publisher    = {OpenReview.net},
  year         = {2024}
}

@article{DBLP:journals/corr/seedvc,
  author       = {Songting Liu},
  title        = {Zero-shot Voice Conversion with Diffusion Transformers},
  journal      = {CoRR},
  volume       = {abs/2411.09943},
  year         = {2024}
}

@article{DBLP:journals/spl/WangCWXW24,
  author       = {Zhichao Wang and Yuanzhe Chen and Xinsheng Wang and Yuping Wang and Lei Xie},
  title        = {StreamVoice+: Evolving Into End-to-End Streaming Zero-Shot Voice Conversion},
  journal      = {{IEEE} Signal Process. Lett.},
  volume       = {31},
  pages        = {3000--3004},
  year         = {2024}
}

@article{DBLP:journals/corr/MeanFlows,
  author       = {Zhengyang Geng and
                  Mingyang Deng and
                  Xingjian Bai and
                  J. Zico Kolter and
                  Kaiming He},
  title        = {Mean Flows for One-step Generative Modeling},
  journal      = {CoRR},
  volume       = {abs/2505.13447},
  year         = {2025}
}

@article{DBLP:journals/corr/abs-2510-08392,
  author       = {Guobin Ma and
                  Jixun Yao and
                  Ziqian Ning and
                  Yuepeng Jiang and
                  Lingxin Xiong and
                  Lei Xie and
                  Pengcheng Zhu},
  title        = {MeanVC: Lightweight and Streaming Zero-Shot Voice Conversion via Mean
                  Flows},
  journal      = {CoRR},
  volume       = {abs/2510.08392},
  year         = {2025}
}

@article{DBLP:journals/corr/Seed-tts,
  author       = {Philip Anastassiou and
                  Jiawei Chen and
                  Jitong Chen and
                  Yuanzhe Chen and
                  Zhuo Chen and
                  Ziyi Chen and
                  Jian Cong and others},
  title        = {Seed-TTS: {A} Family of High-Quality Versatile Speech Generation Models},
  journal      = {CoRR},
  volume       = {abs/2406.02430},
  year         = {2024}
}

@article{jiang2025ref,
  title={REF-VC: Robust, Expressive and Fast Zero-Shot Voice Conversion with Diffusion Transformers},
  author={Jiang, Yuepeng and Ning, Ziqian and Wang, Shuai and Wang, Chengjia and Bi, Mengxiao and Zhu, Pengcheng and Fu, Zhonghua and Xie, Lei},
  journal={CoRR},
  volume={abs/2508.04996},
  year={2025}
}

@article{quamer2026tvtsyn,
  title={TVTSyn: Content-Synchronous Time-Varying Timbre for Streaming Voice Conversion and Anonymization},
  author={Quamer, Waris and Tseng, Mu-Ruei and Nasrallah, Ghady and Gutierrez-Osuna, Ricardo},
  journal={CoRR},
  volume={abs/2602.09389},
  year={2026}
}

@article{DBLP:journals/taslp/SismanYKL21,
  author       = {Berrak Sisman and
                  Junichi Yamagishi and
                  Simon King and
                  Haizhou Li},
  title        = {An Overview of Voice Conversion and Its Challenges: From Statistical
                  Modeling to Deep Learning},
  journal      = {{IEEE} {ACM} Trans. Audio Speech Lang. Process.},
  volume       = {29},
  pages        = {132--157},
  year         = {2021}
}

\end{document}